\documentstyle[prd,tighten,aps]{revtex}
\begin{document}
\draft

%begin wide text
%@twocolumnfalse\endcsname
%\renewcommand{\theequation}{\thesection . \arabic{equation} }
\title{\bf Inflationary cosmology\\
of the extreme cosmic string}

\author{Pedro F. Gonz\'alez-D\'{\i}az}
\address{Centro de F\'{\i}sica ``Miguel Catal\'an'',
Instituto de Matem\'aticas y F\'{\i}sica Fundamental,\\
Consejo Superior de Investigaciones Cient\'{\i}ficas,
Serrano 121, 28006 Madrid (SPAIN)}
\date{May 5, 1998}

\maketitle

\begin{abstract}
Starting with a study of the cosmological solution to the Einstein
equations for the internal spacetime of an extreme supermassive
cosmic string kink, and by evaluating the probability measure for
the formation of such a kink in semiclassical approximation using
a minisuperspace with the appropriate symmetry, we have found a
set of arguments in favour of the claim that the kinked extreme
string can actually be regarded as a
unbounded chain of pairs of Planck-sized
universes. Once one such universe pairs is created along a primordial
phase transition at the Planck scale, it undergoes an endless
process of continuous self-regeneration driven by chaotic inflation
in each of the universes forming the pair.
\end{abstract}

\pacs{PACS number(s): 11.27.+d, 98.80.Cq, 98.80.Hw}

\renewcommand{\theequation}{\arabic{section}.\arabic{equation}}

\section{\bf Introduction}

It has been currently believed that extreme supermassive strings
with linear energy density as large as $\mu=\frac{1}{2G}$ could
not exist because they would drive all the exterior broken-symmetry
phase to collapse into the core, leaving a pure false-vacuum phase
in which the picture of a cosmic string with a core region of
trapped energy is lost [1]. However, it has been recently argued [2]
that the gravity coupling of supermassive strings is so large that
the field-theory defect would also satisfy the conservation laws
of a gravitational topological defect, that is it should become a
cosmic string kink. Then, the picture of a cosmic string is
somehow retained even for the extreme case $G\mu=\frac{1}{2}$, as
the conical singularity is transformed into the apparent singularity
of a cosmological event horizon which turns out to be surrounded by
a shell of the broken-symmetry phase with width $(\sqrt{2}-1)r_{*}
=(\sqrt{2}-1)(8\pi G\epsilon)^{-\frac{1}{2}}$ ($\epsilon$ being the
uniform string density) that prevents the string from disappearing.

What is most interesting about the resulting extreme string kink is
that it meets all the necessary properties to drive an inflationary
process in its core region [2]. On each of its cross-sectional
surfaces, the extreme string kink exactly possesses the symmetry of
a hemispherical section of the de Sitter kink [3], with the radius
of the string exceeding the size of the corresponding cosmological
horizon. A de Sitter-like inflationary process could then be
spontaneously driven in the string core, without any fine tuning
of the initial conditions. It is the main aim of the present work
to explore this rather intriguing possibility by considering both,
the semiclassical approximation to the quantum cosmological model
that results from the minisuperspace that describes a
string whose cross sections satisfy the symmetry of a de Sitter
space, and the physical constraints
that must be imposed to the parameters that define the extreme string.

Our main conclusion is that an extreme string kink drives the creation
of pairs of de Sitter universes. In each of these pairs, the
universes are connected to each other
by a tunnel at their largest surfaces, and inflate
according to the Linde's chaotic inflationary scenario [4]. This model
may quite naturally be accommodated to a process of continually
self-reproducing inflating-universe pairs [4,5], and is made possible by
the fact that a stringy topological defect frozen in the phase
transition of a field theory with large gravity coupling, $G\mu\sim 1$,
is subject to the back reaction of the gravitational kink. This
action would shrink the cross-sectional area of the string, down to
the Planck scale, $r_{*}\sim M_p^{-1}$, with $M_p$ the Planck mass,
and hence implies an initial value
for the field potential $V_0\sim M_p^4$; i.e. just the initial
conditions for chaotic inflation [4].

The paper is organized as follows. Sec. II briefly reviews the
geometry of the extreme string kink and discusses some of its
topological properties.
In Sec. III we extend the discussion on
the geometry of the kinked string and show that it can be
visualized by the embedding of a six-hyperboloid in space
$E^6$. The instantons that can be associated to the Euclidean metrics
of the extreme string kink are dealt with in Sec. IV. It is seen
that the instantons can be created by two different types of
continuation: the usual Wick rotation and the rotation of
the spacelike quantities characterizing the kinked string geometry.
In Sec. V se consider a minisuperspace model that represents the
extreme cosmic string. We use the Hartle-Hawking no boundary [6]
and Vilenkin "tunneling" [7]
proposals to obtain the classical solution, and this is in turn
employed to get a semiclassical probability measure that predicts
inflation only for the case of the tunneling wave function.
Sec. VI deals in some detail with the nature and
properties of the inflationary process that is driven in the string
core. It turns out that Linde's chaotic inflation is the model that
fits best with the estimated parameters of the kinked string. We
close in Sec. VII, with a brief summary of the results, and
some comments
on the considered model in relation with other scenarios for
string-driven inflation, so as on the notion of universe pairs.

\section{\bf The extreme string kink}
\setcounter{equation}{0}

The motivation to consider an extreme cosmic string kink in the
cosmological context is the physical expectation that,
when realized in a given spacetime, the vacuum
manifold, $M$, of the underlying field model with large gravity
coupling should map, through the Einstein equations, into the
associated gravitational manifold, $M_g$. Since the vacuum manifold
for a cosmic string is not simply connected and has, therefore,
nontrivial loops characterized by a group which is isomorphic to
the group of integers (winding numbers) [8], each of these
noncontractible loops would then map into a noncontractible loop
in each of the resulting, mutually
disconnected components, $M^3$, of the gravitational manifold,
$M_g\cong M^{\infty}$, charaterized with an integer topological
charge, $\kappa=0,\pm 1, \pm 2,...$, of a gravitational kink [9]; i.e.:
one would expect that a string (a topological defect in the
vacuum manifold of the field model signaled by the first,
fundamental homotopy group $\pi_{1}(M)$) with the largest
gravity coupling $G\mu=\frac{1}{2}$ should induce the existence
of a kink in the gravitational field (a topological defect moving
about in spacetime which is signaled by the third homotopy group
of the projective sphere, $\pi_{3}(M^3)$). As a result from this
mapping, the gravitational kink would back-react onto the geometry
of the cosmic string, which becomes consequently distorted.

This mapping can occur only for $G\mu=\frac{1}{2}$ in which case
all of the existing spacetime is in the string interior;
otherwise, the gravitational submanifold corresponding to
the exterior broken-symmetry phase has not to possess any
nontrivial loops and, therefore, $M_g$ cannot in general
be divided into disconnected
pieces $M^3$ each with a nontrivial loop, so preventing the
gravitational kink to exist (i.e. the existing exterior region
with a conical singularity of cosmic strings with $G\mu\ll 1$
does not allow the mapping-induced
creation of a compact region supporting the kink.)

In what follows, let us briefly first review the topological
properties of the kinked extreme string, and then comment on some
aspects of its geometry. The general concept of a gravitational
kink can be introduced by starting with the Lorentz metric $g_{ab}$
of a four-dimensional spacetime as given by a map, $P$, from any
connected three-manifold, $\partial${\bf M},
of the spacetime four-manifold, {\bf M}, into the set of
timelike directions in {\bf M} [10]. Metric homotopy can then
be classified by the degree of this map, and the kink number (or
topological charge) of the Lorentz metric, with respect to a
hypersurface $\Sigma$, can be defined by [10]
\[{\rm Kink}(\Sigma;g_{ab})={\rm deg}(P),\]
so that the gravitational kink can be viewed as a measure of how
many times the light cones rotate around as one moves along
hypersurface $\Sigma$.

In the case of the spacetime of an extreme cosmic string, whose
interior geometry can be visualized as that of a sphere when the
corresponding two-metric is embedded in an Euclidean three-sphere [11],
the pair ($\Sigma;g$) will describe a gravitational kink with
topological charge $\kappa=+1$ if
${\rm Kink}(\Sigma;g)=1$. From the above discussion, one may also
visualize the internal geometry of the extreme string by enforcing the
constant-time sections, $\tau=\tau_{0}$, of the interior metric of
the string [2,11]
\begin{equation}
ds^2=\frac{dr^2}{1-\frac{r^2}{r_{*}^2}}+dz^2+r^2d\phi^2,
\end{equation}
where
\begin{equation}
r=r_{*}\sin\frac{\rho}{r_{*}}
\end{equation}
and $-\infty<z<\infty, 0\leq\phi\leq2\pi, 0\leq\rho\leq r_{*}\arccos(1-4G\mu)$,
to be isometrically embedded in the kinked spacetime. The corresponding
cylindrically-symmetric standard, kinked metric is given by [2,12]
\begin{equation}
ds^2=-\cos 2\alpha d\hat{t}^2\mp 2kd\hat{t}dr+dz^2+r^2d\phi^2,
\end{equation}
where the upper/lower sign of the second term corresponds to a
positive/negative topological charge, $k=\pm 1$, depending on which
of the two coordinate patches required for a complete description of
the kink is being considered [2], and $\alpha$ is the tilt angle of
the light cones in the kink, $0\leq\alpha\leq\pi$. The isometric
embedding will hold if in metric (2.3) we have furthermore
\begin{equation}
\cos 2\alpha=1-\frac{r^2}{r_{*}^2}
\end{equation}
and
\begin{equation}
\hat{t}= \tau_{0}-k\int\frac{dr}{\cos 2\alpha}.
\end{equation}

Actually, a gravitational kink depends only on D-1 of the D spacetime
coordinates, and is spherically symmetric on them [9]. However, the
cylindric coordinate $z$ in metric (2.1) and (2.3) is not going to
play any role in the analysis to follow and, therefore, one could
reduce these metrics just to their hemispherical $z$=const. sections.
On the other hand, one can also embed the $z$=const. sections of metric
(2.3) in an Euclidean space and, hence re-express that metric in an
explicit spherically-symmetric form:
\[ds^2=-\cos 2\alpha d\hat{t}^2-kd\hat{t}dr+r_{*}^2d\Omega_{2}^2,\]
where we have specialized to the case of a gravitational kink with
positive topological charge, $\kappa=+1$, $d\Omega_2^2=d\theta^2
+\sin^2\theta d\phi^2$ is the metric on the unit two-sphere, and
we have used Eq. (2.5).

The metric (2.3) can be then regarded as the metric for the
embedding of metric (2.1), and the kinked time $\hat{t}$, as the
corresponding embedding function. Hence, one can obtain an
embedding "rate"
\begin{equation}
\frac{d^2 r}{d\hat{t}^2}=\frac{2r}{r_{*}}\left(\frac{r^2}{r_{*}^2}-1\right),
\end{equation}
which tells us that the embedding surface would flare either outward
if $r<r_{*}$, or inward if $r>r_{*}$. The string metric (2.1) should
now be interpreted as a kinked boundary in the space with kinked
spacetime (2.3).

If the isometric embedding of metric (2.1) in metric (2.3) holds,
from (2.2) and (2.4) we have $\cos^2\theta=\cos 2\alpha$, with
$\theta=\frac{\rho}{r_{*}}$, and if the one-kink is conserved, then
$G\mu$ is enforced to be $\frac{1}{2}$ and $r$ should be
analytically continued beyond $r_{*}$, up to $\sqrt{2}r_{*}$ [2].
This extension creates a spherical shell filled with broken phase at
each $z$-const. section, preventing the extreme string with
$G\mu=\frac{1}{2}$ from disappearing, and converts the conical
singularity at $r=r_{*}$ into a de Sitter-like cosmological
singularity (horizon) [2]. All of the topological charge of the
kink would then be confined within the shell, that is within a
finite compact region beyond the cosmological horizon that
extends up to $r=\sqrt{2}r_{*}$. Inside the horizon all
hypersurfaces $\Sigma$ are everywhere spacelike. Thus, as a
consequence from the back reaction of the gravitational field
of the one-kink, the lost picture of a cosmic string with a
core region of trapped energy would be 
somehow recovered for the extreme
string with $G\mu=\frac{1}{2}$. In Sec. VI it will be argued
that this back reaction would also shift the symmetry-breaking
scale, $\eta$, to a value much larger than that is implied by
$\eta^2\sim\mu$, and such that $r_{*}\sim M_p^{-1}$. What
we call throughout this paper and extreme cosmic string kink
(or in shorter terms, extreme cosmic string or extreme string)
is therefore a cylindrical object with nearly the Planck
thickness, whose cross sections possess the geometrical
structure of the hemispherical section of a de Sitter
spacetime which extends just up to $r=\sqrt{2}r_{*}$,
for a cosmological constant $\Lambda=\frac{3}{r_{*}^2}$.

We have established a consistent and regular embedding of the
extreme string metric in a kinked spacetime whose surfaces
would, according to expression (2.6), flare outward at $\sqrt{2}r_{*}$,
with a maximum "rate"
\[\left.\frac{d^2 r}{d\hat{t}^2}\right|_{r=\sqrt{2}r_{*}}=\frac{2\sqrt{2}}{r_{*}}.\]

To stationary observers at the center of the sphere corresponding to
each surface $z$=const., $\tau$=const., the compact shell containing
all the topological charge of the kink  [13] locally coincides with a finite
region of the exterior of either a de Sitter space when the light
cones rotate away from the observers (positive topological charge), or
the time-reverse to de Sitter space if the observers see light cones
ratating in the opposite direction (negative topological charge). In
the latter case, only the region outside the cosmological horizon
would be accessible to stationary observers.

Topology changes can occur in the compact region [13] of the shell
supporting the kink, but not inside the cosmological horizon. Since
all topologies are allowed to occur in such a compact region, it
would be regarded as an essentially quantum-mechanical bounded
region. Actually, in order to preserve an integer topological
charge for the kink, the description of the spacetime of a kinked
string requires two coordinate patches, with a string spacetime
being fully described in each of these patches. Besides,
continuity of rotation of the light cones implies that the external
boundaries of the compact regions supporting the kink in both
patches be identified to each other and, therefore, a topological
change could be induced in the compact external regions of the
hemispherical sections of the
two de Sitter spaces - one in each patch - 
which would thereby be "bridged" to each other
along a connection at $r=\sqrt{2}r_{*}=(16\pi G\epsilon)^{-1}$.
This is what allows the emergence of tunneling processes between
the cosmological horizons of the two de Sitter spaces (see  Sec. VII.)

Since the physical theories involved in their definition are
all time-symmetric, de Sitter and time-reversed de Sitter spaces
must be physically indistinguishable quantum-mechanically [14].
Therefore, the sign $\mp$ of the second term of the nonstatic metric
(2.3) should be regarded as an unphysical artifact coming from a bad
choice of coordinates, such as it is shown by the fact that metric
(2.3) is still geodesically incomplete at the apparent singularity
at $r=r_{*}$. In fact, it will be seen in the next section 
that the maximally-extended line element
obtained from (2.3) using the Kruskal technique no longer contains
any sign ambiguity other than that is related with the choice of
coordinate patch [2]. Throughout this paper we shall then restrict
ourselves to consider only the extreme cosmic string kink with
positive topological charge.

\section{\bf The spacetime of the extreme cosmic string}
\setcounter{equation}{0}

It was obtained in Ref. [2] that the maximally-extended metric
describing the spacetime of an extreme string kink can be written as
\begin{equation}
ds^2=-\frac{4kr_{*}^2}{(k-UV)^2}dUdV+dz^2+r^2d\phi^2,
\end{equation}
where again $k(=\pm 1)$ labels the two coordinate patches required
to describe a complete one-kink,
\begin{equation}
r_{*}= (8\pi G\epsilon)^{-\frac{1}{2}},
\end{equation}
with $\epsilon$ the uniform string density, out to some cylindrical
radius; $U$ and $V$ are the Kruskal coordinates [2]
\begin{equation}
U=\mp e^{-\frac{k\hat{t}}{r_{*}}}\left(\frac{r_{*}-r}{r+r_{*}}\right),\;\;
V=\pm e^{\frac{k\hat{t}}{r_{*}}},
\end{equation}
in terms of which the radial coordinate can be defined as
\begin{equation}
r=r_{*}\left(\frac{k+UV}{k-UV}\right),
\end{equation}
with the time $\hat{t}$ given by
\[\hat{t}=t
-kr_{*}\sqrt{2\left(1-\frac{r^2}{2r_{*}^2}\right)}\]
\begin{equation}
+\frac{1}{2}kr_{*}\ln\left[\frac{\left(1+
\sqrt{4\left(1-\frac{r^2}{2r_{*}^2}\right)}\right)(r-r_{*})}
{\left(1-
\sqrt{4\left(1-\frac{r^2}{2r_{*}^2}\right)}\right)(r+r_{*})}\right],
\end{equation}
where $t$ is the metrical kinked time which is related to the time
entering the metric of the kinkless cosmic string [2].

As it was already pointed out in Sec. II,
the interesting feature of metric (3.1) is that its $z$-const. sections
coincide exactly with
the metric which describes a hemispherical section of
the de Sitter spacetime kink [3] for a positive cosmological
constant $\Lambda=\frac{3}{r_{*}^2}$. Hence, by suitably redefining
the time entering the metric, we can re-express it as the
geodesically-incomplete metric of a static de Sitter space with
cylindrical symmetry, or as the corresponding Robertson-Walker
metric, which are both induced in the embedding of a six-hyperboloid
\begin{equation}
-(x_{0}^2+x_5^2)+\sum_{a=1}^{4}x_a^2=(1+k)H^{-2}, \;\; H=r_{*}^{-1}
\end{equation}
in $E^6$. In this embedding we choose for the metric of the
hyperboloid
\begin{equation}
ds^2=-(dx_0^2+dx_5^2)+\sum_{a=1}^4 dx_a^2,
\end{equation}
with the topology $R_1\times R_5\times S^4$.

In coordinates $T\in(-\infty,\infty), r\in(0,H^{-1}), \Psi_3\in(-\infty,\infty),
\Psi_2\in(0,\pi)$ and $\Psi_1\in(0,2\pi)$, defined by
\[x_5=H^{-1}\sinh\Psi_3, \;\; x_4=H^{-1}\cosh\Psi_3\]
\[x_3=\sqrt{k}H^{-1}\sin\Psi_2\cos(\sqrt{k}\Psi_1)\]
\begin{equation}
x_2=\sqrt{k}H^{-1}\sin\Psi_2\sin(\sqrt{k}\Psi_1)
\end{equation}
\[x_1=\sqrt{k}H^{-1}\cos\Psi_2\cosh(HT)\]
\[x_0=\sqrt{k}H^{-1}\cos\Psi_2\sinh(HT),\]
we can write the static, geodesically-incomplete metric that
corresponds to the kinked metric (3.1) in the form:
\begin{equation}
ds^2=-k\left(1-H^2r^2\right)dT^2+\frac{kdr^2}{\left(1-H^2r^2\right)}
+dz^2+r^2d\phi^2,
\end{equation}
where we have set $r=H^{-1}\sin\Psi_2$, $\Psi_1=\phi$ and $\Psi_3=Hz$.
Of course, using Kruskal coordinates similar, but not equal to (3.3),
one can recover a maximally-extended metric with the same form as
(3.1) from (3.9).

If we now analytically continue the time $T$ so that $T=i\tau$,
then we obtain an Euclidean metric with signature + + + + in the first
coordinate patch $k=+1$, and a Kleinian metric with signature
- - + + in the second coordinate patch $k=-1$.

It is worth noting that the coordinates defined in (3.8) cover
only the portion of space with $x_1>0$ and
$\left|\sum_{a=2}^3 x_a^2\right|<H^{-2}$; i.e. the region inside
the particle and event horizons of an observer moving on $r=0$,
along the entire axis $z$ of the cosmic string. The event horizon
occurs at the apparent singularity of metric (3.9), at
$r=H^{-1}\equiv r_{*}$.

In coordinates $T'\in(-\infty,\infty)$, $\Psi_{3}\in(-\infty,\infty)$,
$\Psi_2\in(0,\pi)$ and $\Psi_1\in(0,2\pi)$, defined by
\[x_5=H^{-1}\sinh\Psi_3, \;\; x_4=H^{-1}\cosh\Psi_3\]
\[x_3=\sqrt{k}H^{-1}\cosh(HT')\sin\Psi_2\cos(\sqrt{k}\Psi_1)\]
\begin{equation}
x_2=\sqrt{k}H^{-1}\cosh(HT')\sin\Psi_2\sin(\sqrt{k}\Psi_1)
\end{equation}
\[x_1=\sqrt{k}H^{-1}\cos\Psi_2\cosh(HT')\]
\[x_0=\sqrt{k}H^{-1}\sinh(HT'),\]
metric (3.7) becomes the line element describing the geometry of a
homogeneous and cylindrically-isotropic kinked spacetime, that is
\[ds^2=-kdT'^2\]
\begin{equation}
+H^{-2}\left[d\Psi_3^2+\cosh^2(HT')\left(kd\Psi_2^2+\sin^2\Psi_2d\Psi_1^2\right)\right].
\end{equation}

Under the analytical continuation $T'=i\tau '$, metric (3.11) is converted
into again an Euclidean definite positive line element for $k=+1$ and a
Kleinian metric for $k=-1$. The $z$ = const. sections of (3.11)
for $k=+1$
correspond to a three-dimensional Robertson-Walker metric whose
spatial sections have topology $S^2$ with radius
$H^{-2}\cosh(HT')$; i.e. they are de Sitter space in three
dimensions. Spatial sections of (3.11) correspond to the
hyperboloid $R\times S^2$. The coordinates (3.10) cover entirely
this cylindrically-deformed de Sitter space which would first
contract around the entire axis $\Psi_3$ until $T'=0$, and
expand thereafter from that axis.

The hyperboloid described by metric
(3.7) is defined for a constant $(1+k)H^{-2}$
which becomes $2H^{-2}$ in the patch $k=+1$ and vanishes in the
patch $k=-1$. In order to keep the same absolute
value for the constant in
the two patches while preserving the same Euclidean signature, we
should introduce a new analytical continuation for the second
patch such that, instead of the usual Euclidean continuation,
we make $z=-i\Theta$, $H=-i\Pi$, $ds=-id\sigma$ (and hence
$r=-i\rho=-i\Pi^{-1}\sin\psi_{2}$), while changing sign of the
parameter $k$ in (3.8) and (3.10). We then obtain
\begin{equation}
d\sigma^2=-k\left(1-\Pi^2\rho^2\right)dT^2-\frac{kd\rho^2}{\left(1-\Pi^2\rho^2\right)}
+d\Theta^2+\rho^2d\phi^2,
\end{equation}
and
\[d\sigma^2=-kdT'^2\]
\begin{equation}
+d\Theta^2+\Pi^{-2}\cos^2(\Pi T')\left(-kd\Psi_2^2+\sin^2\Psi_2d\Psi_1^2\right),
\end{equation}
respectively. These metrics are now both Euclidean for $k=-1$ and
both Kleinian for $k=+1$.

It would then appear that a positive definite metric can only be
achieved if we make the usual Wick rotation in patch $k=+1$,
and the new continuation, where one rotates $z,H$ and the metric
itseft, in the patch $k=-1$. In the following section we shall
discuss in more detail this choice and the physical reason
supporting it.

\section{\bf Kinked extreme string instantons}
\setcounter{equation}{0}

The Euclidean continuation of the extreme string metric which
contains one-kink should correspond to making the Wick rotation
\begin{equation}
\hat{t}=i\hat{\tau},
\end{equation}
where $\hat{t}$ is the kinky time defined by Eq. (3.5). Using
this continuation, we have [2]
\begin{equation}
d\hat{\tau}=-idt-i\left(\tan 2\alpha-\frac{k}{\cos 2\alpha}\right)dr,
\end{equation}
with $\alpha$ again being the angle of tilt of the light cones on
the hypersurfaces and [2]
\begin{equation}
\sin\alpha=\frac{r}{\sqrt{2}r_{*}}.
\end{equation}

The Euclidean continuation (4.1) gives rise to metrics which are
positive definite only if we choose either the usual continuation
$t=i\tau$ in the coordinate patch $k=+1$, or the new continuation
implying $z=-\Theta, r=-i\rho, r_{*}=-i\rho_{*}$ and $ds=-id\sigma$
in the coordinate patch $k=-1$, which were considered in Sec. III.

In order to investigate the instanton structure of the extreme
string kink, let us first re-write metric (3.1) in the form:
\begin{equation}
ds^2=-k(r+r_{*})^2dUdV+dz^2+r^2d\phi^2,
\end{equation}
where use has been made of (3.4). Introducing then the new variables
$x+y=U$ and $x-y=V$ in (4.4), we get
\begin{equation}
ds^2=-k(r+r_{*})^2(dx^2-dy^2)+dz^2+r^2d\phi^2.
\end{equation}
The following relations will then hold
\begin{equation}
UV=x^2-y^2=k\left(\frac{r-r_{*}}{r+r_{*}}\right)
\end{equation}
\begin{equation}
\frac{U}{V}=\frac{x+y}{x-y}=
ke^{-\frac{2k\hat{t}}{r_{*}}}\left(\frac{r-r_{*}}{r+r_{*}}\right)
\end{equation}

The origin of radial coordinate $r=0$ lies on the surfaces
$y^2-x^2=k$, and the cosmological event horizon $r=r_{*}$ lies
on the surfaces $x^2-y^2=0$, on both coordinate patches. One
can avoid the region either beyond the horizon or inside the
horizon by defining new coordinates, $\zeta=ix$ or $\xi=iy$,
respectively. For the first choice, the metric (4.5) takes the form
\begin{equation}
ds^2=k(r+r_{*})^2(d\zeta^2+dy^2)+dz^2+r^2d\phi^2,
\end{equation}
which is in fact positive definite in the patch $k=+1$, and has
Kleinian signature in the patch $k=-1$. For the choice $\zeta=ix$,
the radial coodinate is defined by
\begin{equation}
\zeta^2+y^2=k\left(\frac{r_{*}-r}{r_{*}+r}\right).
\end{equation}
Then, on the section on which $y$ and $\zeta$ are both real (the
usual Euclidean section for patch $k=+1$ or the Kleinian section for
patch $k=-1$), $\frac{r}{r_{*}}$ will be real and smaller or equal
to 1 on patch $k=+1$, and take on values in the interval
$\sqrt{2}\geq\frac{r}{r_{*}}\geq 1$, on the patch $k=-1$; the
upper limit $\sqrt{2}$ of this interval being imposed by the
continuity of the light-cone tipping on the surfaces at
$\alpha=\frac{\pi}{2}$ [2].

We define now the imaginary time by $t=i\tau$. This continuation
leaves invariant the form of the metric (4.8) and is therefore
compatible with the choice $\zeta=ix$. Then, from Eq. (4.7) we
can obtain
\begin{equation}
y-i\zeta=\sqrt{k}F(r,r_{*})(y^2+\zeta^2)^{\frac{1}{2}}e^{-\frac{ik\tau}{r_{*}}},
\end{equation}
where
\begin{equation}
F(r,r_{*})=\exp\left({\sqrt{2\left(1-\frac{r^2}{2r_{*}^2}\right)}}\right)
\left[\frac{1-2\sqrt{\left(1-\frac{r^2}{2r_{*}^2}\right)}}{1+2\sqrt{\left(1-\frac{r^2}{2r_{*}^2}\right)}}\right]^{\frac{1}{2}}.
\end{equation}

It follows that for this time continuation $\tau$ is periodic with
period $2\pi kr_{*}$, which exactly corresponds to the inverse to the
temperature of the isotropic background of thermal radiation that
is emitted in the space of an extreme string kink [2]. On the
considered Euclidean section, $\tau$ has then the character of
an angular coordinate which rotates about the "axis" $r=0$
clockwise in patch $k=+1$, and anti-clockwise about the "axis"
$r=r_{*}$ in patch $k=-1$.

Since for the spacetime being considered the boundary term in the
action must vanish (see Sec. V), the action of the instanton will
be evaluated using the scalar curvature $R$ only. This action
turns out to be
\begin{equation}
I_{k=\pm 1}=\frac{i\pi M_{p}^2}{2\Lambda},
\end{equation}
where $M_p$ is the Planck mass.

For the second choice of coordinate, $\xi=iy$, the metric (4.5)
takes the form:
\begin{equation}
ds^2=-k(r+r_{*})^2(dx^2+d\xi^2)+dz^2+r^2d\phi^2,
\end{equation}
which is positive definite in patch $k=-1$ and Kleinian in patch
$k=+1$. For this coordinate choice, the radial coordinate becomes
defined by
\begin{equation}
x^2+\xi^2=k\left(\frac{r-r_{*}}{r+r_{*}}\right);
\end{equation}
so, on the section on which $x$ and $\xi$ are both real (the usual
Euclidean section for patch $k=-1$ and the Kleinian section for
patch $k=+1$), the ratio $\frac{r}{r_{*}}$ will now take on values
within the interval $(\sqrt{2},1)$ for patch $k=+1$, and will be
smaller or equal to unity in patch $k=-1$.

Since we have defined the imaginary time $\tau$ for the first
choice $\zeta=ix$, according to Eq. (4.2), we should now define
the imaginary quantities that make the term
$(\sin 2\alpha-k)dr/\cos 2\alpha$ imaginary. Using (4.3) one can
see that such quantities are the imaginary of radial coordinate
$r$ and the imaginary of the extreme string parameter $r_{*}$.
We then define $r=-i\rho$ and $r_{*}=-i\rho_{*}$, while keeping
time $t$ real. In order for this definition to be compatible with
the coordinate transformation $\xi=iy$, one should require that this
definition leaves metric (4.13) formally unchanged, and this
can only be accomplished if coordinate $z$ and the metric itseft,
$ds$, are also continued into their imaginary values, so that
$dz=-id\Theta$ and $ds=-id\sigma$, while keeping the angular
coordinate $\phi$ real, such as it was done in Sec. III.

From Eq. (4.7) we get then
\begin{equation}
x-i\xi=\sqrt{k}F(\rho,\rho_{*})(x^2+\xi^2)^{\frac{1}{2}}e^{-\frac{ikt}{\rho_{*}}}.
\end{equation}
It is now the Lorentzian time $t$ which becomes periodic with
period $2\pi k\rho_{*}$, on the new instantonic section. Therefore,
$t$ would have the character of an angular coordinate on this
section: it will rotate clockwise about the "axis" $\rho=\rho_{*}$
on the coordinate patch $k=+1$, and anti-clockwise about the
"axis" $\rho=0$ on the coordinate patch $k=-1$. The instantonic
action on this new Euclidean section can again be computed from the
scalar curvature only. It is:
\begin{equation}
I_{k\pm 1}=\frac{i\pi M_p^2}{2\Lambda_{E}},
\end{equation}
with $\Lambda_{E}=\frac{3}{\rho_{*}^2}$, on both coordinate patches.

A stationary observer at the origin of the radial coordinate $r$ in the
patch $k=+1$ would interpret the above two instantonic sections
as providing the probability of the occurrence in the vacuum state,
\[P\sim\exp\left(-2I_{k=\pm 1}\right),\]
of an extreme string with positive energy and internal radius
$r_{*}$ on the coordinate patch $k=+1$, and an extreme string with
negative energy and radius $\rho_{*}$ on the coordinate patch
$k=-1$. If the stationary observers were on the patch $k=-1$, then
it would get the same interpretation, but now the positive-energy
string with radius $r_{*}$ would occur in patch $k=-1$ and the
negative-energy string with radius $\rho_{*}$ in patch $k=+1$,
provided that the topological charge of the kink continues being
positive with respect to the observer. In both cases, the
spacetimes of the two strings should join to each other on the
surfaces $\sqrt{2}r_{*}$ (or $\sqrt{2}\rho_{*}$), beyond their
horizons. The resulting whole geometrical construct could then
be regarded as a kinked extreme string pair.

\section{\bf Creation of universe pairs}
\setcounter{equation}{0}

In this section we consider the vacuum solution of the Euclidean
Einstein equations with a cosmological constant
$\Lambda=\frac{3}{r_{*}^2}$, describing the interior of a kinked
cylindric extreme cosmic string. Assuming that the no boundary
condition [6] is satisfied at the initial time, the Euclidean
action for the system can be written
\[I_{E}=-\frac{M_p^2}{16\pi}\int_{M}d^4 x\sqrt{g}(R-2\Lambda)\]
\begin{equation}
-\frac{M_p^2}{8\pi}\left(\int_{\partial M}d^3 x\sqrt{h}K
-\int_{\tau=0}d^3 x\sqrt{h}K\right),
\end{equation}
where $g$ and $h$ refer to the four- and three-metric, respectively,
$R$ is the scalar curvature, and $K$ is the trace of the second
fundamental form, both on the chosen boundary $\partial M$ and at
the initial time $\tau=0$. The latter term must be added because
it is an essential prescription of the no boundary proposal [6]
that there should be no boundary at the initial time $\tau=0$,
so this term explicitly adds the contribution from $\tau=0$
back in. Such a term could, in principle, be nonzero for the
kind of topologies we are going to use. If we introduce any other
boundary conditions, the last term in (5.1) should not be added.

According to the characteristics of the spacetime dealt with in
the previous sections, we choose a minisuperspace model given by
the Euclidean metric
\begin{equation}
ds^2=kd\tau^2+dz^2+b^2\left(kd\Psi_{2}^2+\sin^2\Psi_2 d\Psi_1^2\right),
\end{equation}
with $b\equiv b(\tau)$ the scale factor.

From (5.2) we obtain for the Ricci-scalar
\begin{equation}
R=-k\left(3\frac{\dot{b}^2}{b^2}+4\frac{\ddot{b}}{b}-\frac{1}{b^2}\right),
\end{equation}
where the dot means time derivative, $^{.}=\frac{d}{d\tau}$.
In order to derive an expression for the Euclidean action in our
minisuperspace model, we inscribe each spatial section $S^2$ of the
metric in a cylinder so that the origin of the angular
coordinate $\Psi_2$ lies on the $z$-axis of the cylinder.
If, for a two-sphere located at the origin of the cylindrical
coordinates, we choose as the metric the Euclidean continuation
of (3.11), then we should take $z=r\cos\Psi_2$, since $\Psi_1=\phi$
and $r=H^{-1}\cos(H\tau)$, with $H=\sqrt{\Lambda}$. It follows
that in the Euclidean sector that corresponds to just one inscribed
sphere, $-H^{-1}\leq z\leq H^{-1}$, and hence we obtain for the
action on that sector in the two coordinate patches
\begin{equation}
I_{E}=-\frac{M_p^2}{H}\left(\int Nd\tau\left(\frac{\dot{b}^2}{N^2}
+1-\Lambda b^2\right)+\left[\frac{b\dot{b}}{N}\right]_{\tau=0}\right),
\end{equation}
where $N$ is the lapse function and the second term corresponds to
the last surface term of Eq. (5.1).

In the gauge $N=1$,
the equation of motion for $b$ and the Hamiltonian constraint will
be:
\begin{equation}
\ddot{b}=-b\Lambda
\end{equation}
\begin{equation}
1-\dot{b}^2-\Lambda b^2=0.
\end{equation}
A solution to these equations is
\begin{equation}
b(\tau)=H^{-1}\sin(H\tau),
\end{equation}
which looks like Nariai spacetime [15]. This solution naturally
satisfies the no boundary condition as, at $\tau=0$
\begin{equation}
b=0,\;\;\; \dot{b}=1.
\end{equation}

Note that the boundary term in Eq. (5.4) vanishes for this solution.
Now, if we choose a path along the ${\rm Re}\tau$ axis from 0 to
$\frac{\pi}{2H}$ [16], the solution (5.7) will describe twice half
of the Euclidean $R\times S^3$ instanton, each time in a
coordinate patch. If the path is continued from
${\rm Re}\tau=\frac{\pi}{2H}$ parallel to the imaginary axis
${\rm Im}\tau$ [16], then $b$ will still remain real, with
\begin{equation}
\left.b({\rm Im}\tau)\right|_{{\rm Re}\tau=\frac{\pi}{2H}}
=H^{-1}\cosh(H{\rm Im}\tau).
\end{equation}

The scale factor (5.9) describes twice half of a Lorentzian universe,
each time on a coordinate patch. The spacetime sections of this
universe can be visualized as being formed by two spheres (one
in each patch) growing up from the original size of the extreme
string kink. The physical interpretation would be that of a pair
of physical universes spontaneously created out from the extreme
string kink, the two universes in each pair being formed at
the same time, and joined to each other at the surfaces
$\sqrt{2}b$, beyond their respective observable regions of radius
$b$.
(Note that for $G\mu=\frac{1}{2}$ there is no exterior
space to the string shell in one coordinate patch, except that
of the string shell in the other coordinate patch.) Thus, the
observable parts of the two universes inside the cosmological
horizon accelerate away from
each other as $b$, and hence their mutual separation
$2(\sqrt{2}-1)b$, grows.

On the other hand, since for the considered solution the second
term of (5.4) vanishes, the real part of the action comes
enterely from the first term. Besides, the Lorentzian segment
of the chosen path only contributes to ${\rm Im}I_E$, so we
obtain finally for the real part of the action corresponding to
the two universes in a pair
\begin{equation}
{\rm Re}I_E=-\frac{2M_p^2}{H}\int_0^{\frac{\pi}{2H}}d\tau\cos^2 H\tau
=-\frac{\pi M_p^2}{2\Lambda},
\end{equation}
and the semiclassical probability measure for creation of a pair
of such universes will be given by
\begin{equation}
P_{HH}\sim\exp\left(\frac{\pi M_p^2}{\Lambda}\right).
\end{equation}
One should compare (5.11) with the corresponding probability for
a single spherically-symmetric de Sitter universe satisfying the
no boundary proposal [16]
\begin{equation}
P_{HH}^{dS}\sim\exp\left(\frac{3\pi M_p^2}{\Lambda}\right).
\end{equation}
One can conclude that, if the no boundary holds, then the
creation of a single de Sitter universe is by far a more
probable process than the creation of a universe pair
according to the mechanism considered in this work. On the other
hand,
larger values of the matter field $\varphi$ entering the potential
\[V=\frac{\Lambda M_p^2}{8\pi},\]
with $\Lambda=2\pi G\lambda\varphi^4$,
will be strongly suppressed according to (5.11). Therefore, also
when applied to a universe pair created out from an extreme string kink,
the no boundary proposal would prevent inflation to occur
in each universe of the pair. It is worth noting that the main
objection raised by Linde [17] and Vilenkin [7] against the no
boundary proposal is that it is unable to predict inflation. This
result was to be expected as the universe pairs created from an
extreme string kink with finite size cannot be self-contained,
but is a consequence from the existence of a previous physical
spacetime reality that necessarily includes
some phase transition.

It appears most appropriate that the boundary conditions for
these universe pairs be related with a "tunneling" condition [7],
where the wave function $\psi(b)$ does not vanish for $b=0$. The
reason is that in the case under consideration, all of the space
is confined within the finite interior region of the kinked
extreme string, and thereby all three-geometries, defined on
the hypersurfaces, of the corresponding superspace should map
onto the same compact region. Now, since the kink makes the
light cones to continuously rotate on the hypersurfaces, the
time direction would rotate as well,
pointing always toward the
superspace's boundary and covering all possible directions only
if the two coordinate patches and both, positive and negative
gravitational topological charges are used. Actually, the time
direction of the light cones point toward the interior of
superspace in the second coordinate patch, $k=-1$ [3]. However,
because the energy of the modes becomes negative in this patch
[2], their time direction will be equivalent to that of a
positive-energy flux for outgoing modes, just as in the first
coordinate patch, $k=+1$.
Therefore, the wave functional
representing the quantum state of the kinked extreme string
should include only all the outgoing modes carrying positive
flux out of superspace, and this just expresses the Vilenkin's
tunneling boundary condition [7]; i.e.:
the quantum creation of the
extreme string from nothing, meaning by "nothing"
the ausence of an exterior
spacetime.

The Wheeler-DeWitt equation for $\psi(b)$ can be written
\begin{equation}
\left\{x^{-p}\frac{\partial}{\partial x}x^p\frac{\partial}{\partial x}
+\left[\left(\frac{M_p}{H}\right)^2-\frac{1}{4}x^2\right]\right\}\psi(x)=0,
\end{equation}
where $x=2M_p b$ and the parameter $p$ represents the factor-ordering
ambiguity. For the choice $p=0$ and the general boundary condition
$\psi(x)\rightarrow 0$, as $b\rightarrow\infty$, we obtain the
general solution
\begin{equation}
\psi(b)=D_{\frac{M_p}{H}-\frac{1}{2}}(2M_p b),
\end{equation}
with $D$ being the parabolic cylinder function for $b>0$.

However, for the purposes of this work, it suffices obtaining the WKB
solutions to (5.13). In the most natural case that corresponds
to a tunneling boundary condition, the real part of the
underbarrier ($b<H^{-1}$) solution to (5.13) has the form [7]
\begin{equation}
\psi_{T}\sim\psi_{HH}^{-1}\sim\exp\left(-\frac{\pi M_p^2}{2\Lambda}\right),
\end{equation}
where $\psi_{HH}$ is the (underbarrier) WKB solution satisfying the
no boundary condition and we have disregarded the pre-exponential
factor. It follows that the tunneling semiclassical probability measure for
the creation of a pair of universes must be given by
\begin{equation}
P_T\sim\exp\left(-\frac{\pi M_p^2}{\Lambda}\right),
\end{equation}
for which larger values of the field $\varphi$ will be strongly
favoured, rather than suppressed. Moreover, by comparing (5.16)
with the corresponding probability for a single de Sitter
universe satisfying the tunneling proposal, obtained from
(5.12) by the same approximate procedure, that is
\begin{equation}
P_{T}^{dS}\sim\exp\left(\frac{-3\pi M_p^2}{\Lambda}\right),
\end{equation}
we deduce that, in this case the process of pair creation turns
out to be by far more likely than the creation of a single de
Sitter universe.

Thus, if one insists in an inflationary cosmological model in
which the universes are created by the mechanism of pair formation
considered in this work, it appears that one should choose as
initial condition an original tunneling from nothing, where the
outgoing modes of the wave function point toward the complete
singular boundary of superspace only if two coordinate patches
and both, positive and negative topological charges
are used to describe the kinked string.

\section{\bf Inflation in universe pairs}
\setcounter{equation}{0}

There are two main reasons in favour of the idea that, once
simultaneously created, the two universes forming a pair 
along the cosmic string kink inmediately
undergo a separate, but equivalent
inflationary process. First of all, it is
the fact that the radius of a spherical extreme string kink
exceeds the size of its corresponding cosmological horizon.
This is a straightforward consequence from the de Sitter
structure of the $z$=const. sections of the
extreme-string kink interior, which should
induce it to quite naturally drive a exponential expansion, without
fine tuning of the initial conditions. This implication agrees
with the proposal by Linde and Linde [18] and
Vilenkin [19] that inflation may be
generated in the core of topological defects for sufficiently large
gravity coupling. The second reason for an inflationary process
in each of the spacetimes of the pair has been discussed at the
end of Sec. V; i.e.: the quantum-cosmological prediction of an
increase of the semiclassical probability measure as the involved
matter field $\varphi$ becomes larger, provided that the initial
state satisfies a tunneling boundary condition. It is well-known
that inflation can only be driven for high initial values of the
matter field [17].

In order for the event horizon at $r=r_{*}$ of an extreme string kink
(which has linear energy density $\mu=\frac{1}{2}M_p ^2$ and
possesses $z$=const. sections with
the symmetry of the hemispherical section of a de
Sitter kink [2,3]) to be a cosmological horizon with size
\[H^{-1}=r_{*}=\left(\frac{3}{8\pi GV_0}\right)^{\frac{1}{2}},\]
one must choose the uniform string density to be $\epsilon=
\frac{V_0}{3}$, so that the cosmological constant becomes
$\Lambda=8\pi GV_0=\frac{3}{r_{*}^2}$. The crucial point now
is that, if we use the approximate expression [20] $\mu\sim\eta^2$,
where $\eta\sim M_p$ is the symmetry-breaking scale of the
underlying model
\begin{equation}
V(\varphi)=\frac{1}{4}\lambda(\varphi_a\varphi_a-\eta^2)^2,
\;\; a=1,2 ,
\end{equation}
then $V_0=\frac{1}{4}\lambda\eta^4\sim\lambda M_p^4$ and, since
$\lambda\ll 1$, $V_0\ll M_p^4$ and
$r_{*}=(8\pi G\epsilon)^{-\frac{1}{2}}\sim\lambda^{-\frac{1}{2}}M_p^{-1}
\gg M_p^{-1}$.

In this case, the initial conditions for inflation
would correspond to the following set of field parameters:
\begin{equation}
\partial_{\mu}\varphi\partial^{\mu}\varphi\ll M_p^{4},\;\;
V_0(\varphi_0)\ll M_p^{4},\;\; R^2\ll M_p^4,
\end{equation}
where $R$ is the Ricci-scalar. Under these conditions, the only
region accessible to stationary observers is that with radius
$r_{*}=H^{-1}\gg M_p^{-1}$ and, in order for $H^{-1}$ to recede
from $r_{*}$ slowly enough for any possible particles and
other inhomogeneities to not have any effects on events taking
place inside the horizon, we should have $\dot{H}\ll H^2\sim r_{*}^{-2}$.
On the other hand, if the initial values of the potential and
scalar field are $V_0\ll M_p^4$ and $\varphi_0\sim\eta\sim M_p$,
respectively, in a region of size $\ell\sim H^{-1}\gg M_p^{-1}$,
the variation of the field $\varphi$ would be
$\bigtriangleup\varphi\ll\eta\sim\varphi_0\sim M_p$. This would
mean that the given region will be largely homogeneous and isotropic,
and therefore describable as a Friedmann spacetime, where
\begin{equation}
H^2+\frac{s}{b^2}=\frac{\dot{b}^2}{b^2}+\frac{s}{b^2}
=\frac{8\pi}{6M_p^2}\left(\dot{\varphi}^2+\nabla\varphi^2
+2V\right)
\end{equation}
\begin{equation}
\ddot{b}+3\frac{\dot{b}\dot{\varphi}}{b}-\frac{\nabla\varphi}{b^2}
=-\frac{dV}{d\varphi},
\end{equation}
where $s$ is the spatial curvature constant.

However, even though for a sufficiently uniform field $\varphi$ we still
may have $(\nabla\varphi)^2\ll V$ and, if an exponential expansion
is assumed, $\dot{b}^2\gg 1$ and $\ddot{\varphi}\ll\frac{dV}{d\varphi}$,
so that
\begin{equation}
\frac{\dot{b}^2}{b^2}\simeq\frac{8\pi}{6M_p^2}\left(\dot{\varphi}^2
+2V\right)
\end{equation}
\begin{equation}
3H\dot{\varphi}\simeq -\frac{dV}{d\varphi},
\end{equation}
we now had from (6.6)
\begin{equation}
\dot{\varphi}^2\simeq\frac{M_p^2}{24\pi V}\left(\frac{dV}{d\varphi}\right)^2
\sim\frac{\lambda M_p^2\varphi^2}{6\pi},
\end{equation}
which, for $\varphi_0\sim M_p$, becomes $\dot{\varphi}^2\sim V$
and the term with $\dot{\varphi}^2$ could not be dropped off from
Eq. (6.5). It follows that with the set of values used for the
parameters of our model given by (6.2),
neither Eq. (6.5) can give rise to a de
Sitter exponential expansion of the scale factor,
$b\sim b_0\exp(Ht)$, nor
$\dot{\varphi}^2\sim V$ would imply by itself a stress-tensor
$T_{\mu\nu}\sim g_{\mu\nu}V$, predicting a de Sitter state
equation $p\sim-\rho$, with $p$ the pressure and $\rho$ the
energy density.

This result is clearly contradictory with the fact that we
started with a de Sitter space underlying an inflationary process.
Any possible way out of this inconsistency would require having
$V_0\sim M_p^4$, instead of $V_0\sim\lambda M_p^4$, with
$\lambda\ll 1$. Actually, the crucial point which decides on
these two initial values of the potential of the model is the
relation between the symmetry-breaking scale $\eta=\varphi_0$
and the string mass per unit length $\mu\sim M_p^2$. Choosing
$\eta^2\sim\mu$, as usual, leads to the above inconsistency,
but if we assume a relation $\eta^2\gg\mu$, i.e. if
$\varphi_0\gg M_p$, then we see from the Friedmann equations
(6.3) and (6.4) that $\dot{\varphi}^2$ must be much smaller
than $V$, and hence
\[\frac{\dot{b}^2}{b^2}\simeq\frac{8\pi V}{3M_p^2},\]
so that we finally recover the wanted expressions for the scale
factor, $b\sim b_0\exp(Ht)$, and the state equation,
$p\sim -\rho$, required to keep full
consistency in the inflationary model. Besides, if $\eta^2\gg\mu$
and $V\sim M_p^4$, we have
\begin{equation}
r_{*}\sim\mu^{-\frac{1}{2}}\sim M_p^{-1},\;\; \Lambda\sim M_p^2.
\end{equation}

But, why should one use the inequality $\eta^2\gg\mu$ instead of
the approximate relation $\eta^2\sim\mu$ of current cosmic string
theory [20]?. First of all, we remind that the concept of linear energy
density (or mass per unit length) $\mu$, so as the approximate
relation $\eta^2\sim\mu$ are not unambiguously defined and that
they can only be approximately applied to cosmic strings with
moderate tension at symmetry-breaking scales quite lower than
the Planck scale [20]. More importantly, making a cosmic string
obey the conservation laws of the one-kink gravitational defect
changes the internal string structure because of the
gravitational back-reaction from the spacetime kink.

One would expect that the
characteristic radial coordinate $r$ of the string with
$G\mu=\frac{1}{2}$ classically collapses to a point (Ref. Eq.
(2.2)) when the kink is not present, but it would stop
shrinking at a minimum, nonzero value of the order the
Compton wavelength corresponding to the symmetry-breaking scale
$r\sim\delta\varphi_{\eta}\sim\eta^{-1}\sim M_p^{-1}$ (the
maximum energy-scale of the theory) when the quantum
structure of the spacetime kink is considered. The observable
minimum value of $r$ in the kinked extreme string cannot
correspond to the Compton wavelength of the Higgs boson since
all of the observable region of the existing spacetime is
filled with false vacuum, with no trace of the broken-symmetry
phase, so for stationary observers at $r=0$ there will be not
symmetry breaking and, therefore,
the Higgs bosons would not exist.

The parameter $\mu$ is usually defined [20] as a linear energy
density, such that
\[\mu\sim V_0(\delta\varphi_A)^2,\]
where $V_0\sim\lambda\eta^4$ and
$\delta\varphi_A\sim m_{\varphi}^{-1}=\lambda^{-\frac{1}{2}}\eta^{-1}$
is the Compton wavelength of the Higgs boson, so $\mu\sim\eta^2$ holds
only for the situations in which the Higgs boson
is observable; i.e. for cosmic strings with $G\mu\ll 1$, which
are able to have external observers. For strings with $G\mu=\frac{1}{2}$
having only internal stationary observers living in the false vacuum,
$\mu$ should instead be defined as
\[\mu\sim V_0(\delta\varphi_{\eta})^2\sim\frac{m_{\varphi}}{\delta\varphi_{A}}
\sim\lambda\eta^2,\]
so, for $\mu\sim M_p^2$ and $\lambda\ll 1$, we in fact obtain
\begin{equation}
\eta^2=\varphi_0^2\gg\mu\sim M_p^2,\;\;
\partial_{\mu}\varphi\partial^{\mu}\varphi,V_0,R^2\sim M_p^4,
\end{equation}
as the initial conditions for the inflationary process in the
kinked extreme string.

It is worth noting that the inflationary scenario that results
from the above initial conditions is just that of Linde's
chaotic inflation [4], as applied separately to each of the
universes in a pair. For each universe, we then have
\begin{equation}
b(t)\simeq b_0\exp\left(\frac{t}{r_{*}}\right)
\end{equation}
\begin{equation}
\varphi(t)\simeq\eta\exp\left(-\sqrt{\frac{\lambda}{6\pi}}\frac{t}{r_{*}}\right),
\end{equation}
so that
\begin{equation}
b\simeq b_0\exp\left[\sqrt{\frac{2\pi\lambda}{3M_p^2}}(\eta^2-\varphi^2)\right].
\end{equation}
Like in chaotic inflation, here the inflationary regime would end
when the field $\varphi\leq r_{*}\sim M_p$. Besides, since
$\varphi_0=\eta\gg M_p$, the overall inflation factor is
\begin{equation}
E\sim\exp\left[\sqrt{\frac{2\pi\lambda}{3M_p^2}}\eta^2\right].
\end{equation}

One may conclude that most of the physical volume of each pair of
universes created out of an kinked extreme string comes into
being as a result from the inflation of regions with a size
$\ell$ which, in our case, are confined in a finite interval
$r_{*}=H^{-1}<\ell\leq\sqrt{2}H^{-1}$, and were initially
filled with a sufficiently homogeneous and slowly varying (i.e.
$\bigtriangleup\varphi\sim M_p\ll\eta$), extremely large field
$\varphi=\eta\gg M_p$.

In order to have sufficiently homogeneous and isotropic universes,
all the particles and other inhomogeneities initially present
within a sphere of radius $H^{-1}$ should cross the horizon by
a time $H^{-1}$, so that they will have no effect on events
taking place inside the horizon (no hair theorem [21]). In our
case, however, the exterior region is restricted to extend
only up to the surface at $\sqrt{2}H^{-1}$, and this
boundary surface in a coordinate patch
is identified with the similar surface of
the exterior region of the other universe in the other
coordinate patch. Here the no hair theorem is ensured to
hold by the mutual annihilation of particles and inhomogeneities
coming from the two universes in each pair [2].

Because the parameters of our inflationary model have the values
required by chaotic inflation, the global geometry in the two
coordinate patches of the infinite string will show remarkable
inhomogeneities on the largest scales. One obtains that regions with
size $H^{-1}$ will grow by a factor $e$ in a typical time $H^{-1}$
and then subdivide into $\sim e^3$ regions of the same size as
the original one, each with a field $\varphi$ which differs
from the original one by
\[\delta\varphi\sim\sqrt{\frac{\lambda}{6\pi}}\frac{\varphi^2}{M_p}\]
due to quantum fluctuations with wavelength $\geq H^{-1}$ [4].
Nearly half of these new regions will have fields $\varphi$ which are
larger than $\varphi_0$ by $\delta\varphi$ and hence $V\gg M_p^4$,
so inflation would be cut short on them. This process would repeat
during the next time interval $H^{-1}$ in the two coordinate
patches, and so on, to finally reproduce the picture of an infinite
number of self-regenerating pairs of inflationary universes,
where no future singularity is needed or possible [4]. On the
other hand, since the wave function for each of these regions
should satisfy a tunneling initial condition such that the present
approach becomes based on a picture where each pair of
universes nucleates from an isolated gravitational topological defect
in the ausence of any exterior spacetime ("nothing"),
no past singularity seems to have existed or be needed either.

\section{\bf Conclusions and further comments}
\setcounter{equation}{0}

This paper develops and somewhat extends the idea suggested in
Ref. [2] that an extreme cosmic string with linear energy
density $\mu=\frac{1}{2G}$ must exist as a result of its
gravitational kinked structure, and be the seed for an
inflationary process.

We began by considering in detail some necessary geometrical
aspects of the extreme string kink, and then worked out the
structure of the associated instantons. There are two instantonic
sections for this kink, one which is achieved by applying the
customary Wick rotation on time coordinate, gives a positive
definite kinked Kruskal metric only in the first of the two
coordinate patches needed for a complete description of the
one-kink, and the other, arrived at by analytically continuing on the
spacelike coordinates, which produces an Euclidean metric only
on the second of these patches. These instantons can be continued
from one another by glueing the string metrics of the two
patches at their maximum surfaces. This is a requirement from
continuity and completeness on the kink, and therefore, the
glued instantons represent the probability of creation in
the vacuum state of a pair of extreme strings. It has been
also shown that the spacetimes on the two coordinate patches
must inflate separately.

The inflationary process driven in the kinked extreme string
core has been considered under two different points of view.
By using the machinary of the semiclassical approximation to
quantum cosmology, one can see that inflation can only be driven
if one imposes a tunneling boundary condition,
and by discussing Einstein
equations, we achieve the result that kinked extreme strings
undergo a chaotic inflationary process along their core.

We argued that the effect of having a conserved
gravitational topological charge in the high energy string
is two-fold. On the one hand, it induces the creation of a
protecting shell of true vacuum surrounding the string core
and, on the other hand, it gives rise to a gravitational
back-reaction that fits the string size to be of the order
the Planck scale. Whether or not our universe is in a pair
created according to the mechanism suggested in this work
would be a matter requiring further investigation.

The inflationary scenario proposed in this paper can be
thought of as a natural process as far as a phase transition
occurred at the Planck time, breaking the unification
between gravity and the other forces. If such a transition
took place rapidly enough, then topological defects like
cosmic strings with the Planck tension had been formed
satisfying the conservation laws of a gravitational kink
and having the natural conditions for chaotic inflation
to to be driven in the kinked string core.

The possibility of string-driven inflation was first suggested
by Turok [22] who considered the effects of quantum fluctuations
on a string network in de Sitter space. He obtained that strings
able to produce the wanted effect had to be of the Planck scale.
The main criticism that can be raised against models of
string-driven inflation is that Planck-energy strings have
a deficit angle $\bigtriangleup=8\pi G\mu$ (even exceeding $2\pi$)
which would cause strong deviations from homogeneity and isotropy
on the horizon scale [20], and so the Friedmann equations will not
be applicable in this case. Actually, for the kinked string discussed
in this work, there is not anything like a deficit angle. Due to
the gravitational back-reaction caused by the conservation of
the gravitational topological charge, the geometry of the string
becomes no longer conical, but converts into that of a section of
the de Sitter space, with the conical singularity replaced for
the apparent singularity of a cosmological horizon. Clearly, the
picture of a deficit angle is lost, so one should not expect
the kinked string to induce any deviations from homogeneity
or isotropy in any region of the inflating string interior.

Because the kinked extreme string should be regarded as a quantized
geometrical construct [2], it would admit the definition of a maximum
Hagedorn temperature of the form $T_H\sim\sqrt{\mu}\sim M_p$.
It turns out that this temperature is of the same order as the
Gibbons-Hawking temperature $T_{GH}=\frac{1}{2\pi r_{*}}\sim M_p$,
evaluated in Ref. [2] for kinked extreme strings. The same
approximate relation, $T_H\sim T_{GH}$, was assumed by
Aharonov, Englart and Orloff [23] as the basis for suggesting the
possibility of string-driven inflation.

On the other hand, the description of the instantons
associated with the kinked
spacetimes require two coordinate patches. Generally, these instantons
can be regarded as pairs of the corresponding geometrical construct,
one in each patch. This interpretation has been recently discussed
in the context of neutral black-hole pairs [24]. In the case of
the inflating universe pairs considered in this work, the identification
of the two patches only at a tilt angle $\alpha=\frac{\pi}{2}$
leads to an interconnection between the two universes taking
place at the largest, albeit finite surface beyond the
region which is causally connected to stationary observers,
with the universe in one patch being the anti-universe to
the universe in the other patch. This would give rise to an
essentially unobservable bridge between two different
cosmological horizons.

\acknowledgements

\noindent The author thanks Stephen Hawking for hospitality at
the Department of Applied Mathematics and Theoretical Physics,
Cambridge University, UK, where part of this work has been
done. This investigation has been supported by DGICYT under
Research Project No. PB94-0107.


\begin{references}

\bibitem {1} P. Laguna and D. Garfinkle, Phys. Rev. D40, 1011 (1989);
M.E. Ortiz, Phys. Rev. D43, 2521 (1991).
\bibitem {2} P.F. Gonz\'alez-D\'{\i}az, Phys. Rev. D52, 5698 (1995).
\bibitem {3} K.A. Dunn, T.A. Harriot, and J.G. Williams, J. Math.
Phys. 35, 4145 (1994).
\bibitem {4} A.D. Linde, JETP Lett. 38, 149 (1983); Phys. Lett.
129B, 177 (1983); {\it Particle Physics and Inflationary Cosmology}
(Harwood Academic Publishers, London, England, 1990).
\bibitem {5} A.S. Goncharov and A.D. Linde, Sov. Phys. JETP 65, 635 (1987);
A.S. Goncharov, A.D. Linde, and V.F. Mukhamov, Int. J. Mod. Phys.
2A, 561 (1987).
\bibitem {6} J.B. Hartle and S.W. Hawking, Phys. Rev. D28, 2960 (1983).
\bibitem {7} A.D. Linde, Sov. Phys. JETP 60, 211 (1984);
Lett. Nuovo Cim. 39, 401 (1984); A. Vilenkin, Phys. Rev. D30, 549 (1984);
{\it ibid} D33, 3560 (1986); {\it ibid} D37, 888 (1988).
\bibitem {8} T.W.B. Kibble, J. Phys. A9, 1387 (1976).
\bibitem {9} D. Finkelstein and C.W. Misner, Ann. Phys. (N.Y.) 6, 230 (1959).
\bibitem {10} G.W. Gibbons and S.W. Hawking, Phys. Rev. Lett. 69, 1719 (1992).
\bibitem {11} J.R. Gott, Astrophys. J. 288, 422 (1985);
W.A. Hiscock, Phys. Rev. D31, 3288 (1985).
\bibitem {12} D. Finkelstein and G. McCollum, J. Math. Phys. 16, 2250 (1975).
\bibitem {13} A. Chamblin, {\it Kinks and Singularities}, DAMTP Preprint
R95/44.
\bibitem {14} S.W. Hawking, Phys. Rev. D13, 191 (1976); D14, 2460 (1976).
\bibitem {15} H. Nariai, Sci. Rep. Tohoku Univ. 35, 62 (1951).
\bibitem {16} R. Bousso and S.W. Hawking, Phys. Rev. 52D, 5659 (1995).
\bibitem {17} A.D. Linde, {\it Inflation and Quantum Cosmology}
(Academic Press, Boston, 1990).
\bibitem {18} A.D. Linde and D.A. Linde, Phys. Rev. D50, 2456 (1994).
\bibitem {19} A. Vilenkin, Phys. Rev. Lett. 72, 3137 (1994)
\bibitem {20} A. Vilenkin and E.P.S. Shellard, {\it Cosmic Strings and
Other Topological Defects} (Cambridge Univ. Press, Cambridge, England,
1994).
\bibitem {21} G.W. Gibbons and S.W. Hawking, Phys. Rev. D15, 2738 (1977).
\bibitem {22} N. Turok, Phys. Rev. Lett. 60, 549 (1988).
\bibitem {23} A. Aharonov, F. Englert, and J. Orloff, Phys. Lett. 199B,
366 (1987).
\bibitem {24} P.F. Gonz\'alez-D\'{\i}az, in: {\it Theories of Fundamental
Interactions; Maynooth Bicentenary Volume}, ed. T. Tchrakian
(World Scientific, Singapore, 1995); Grav. Cosm. 2, 122 (1996).


\end{references}
\end{document}